\documentclass[aps,prl,preprint,superscriptaddress,longbibliography,floatfix]{revtex4-1}
\usepackage{siunitx}
\usepackage{amsfonts}
\usepackage{amssymb}
\usepackage{amsmath}
\usepackage{upgreek}
\usepackage{color}
\usepackage{graphicx}
\usepackage{hyperref}
\hypersetup{colorlinks,breaklinks,
            urlcolor=blue,
            linkcolor=blue,
            citecolor=blue}

\newcommand{\UCBChem}{Department of Chemistry, University of California, Berkeley, California 94720, USA.}
\newcommand{\LBNL}{Materials Sciences Division, Lawrence Berkeley National Laboratory, Berkeley, California 94720, USA.}
\newcommand{\LBNLC}{Chemical Sciences Division, Lawrence Berkeley National Laboratory, Berkeley, California 94720, USA.}
\newcommand{\JCAP}{Joint Center for Artificial Photosynthesis, Lawrence Berkeley National Laboratory, Berkeley, California 94720, USA.}
\newcommand{\ATLAS}{ATLAS Materials Science Laboratory, Department of NanoEngineering and Chemical Engineering, University of California, San Diego, La Jolla, California, 92023, USA.}
\newcommand{\UCSDMSE}{Materials Science and Engineering, University of California San Diego, La Jolla, California, 92023, USA.}
\newcommand{\UCSDSPEC}{Sustainable Power and Energy Center, University of California San Diego, La Jolla, California, 92023, USA.}
\newcommand{\UoF}{University of Florida, Gainesville, FL 32611, USA.}
\newcommand{\ANL}{Advanced Photon Source, Argonne National Laboratory, Argonne, IL 60439, USA.}
\newcommand{\UoT}{Institute for Solid State Physics, The University of Tokyo, Kashiwa, Chiba 277-8581, Japan.}
\newcommand{\UoTQ}{Trans-scale Quantum Science Institute, The University of Tokyo, Bunkyo-ku, Tokyo 113-0033, Japan.}
\newcommand{\Spr}{RIKEN SPring-8 Center, Sayo, Hyogo, 679-5148, Japan.}
\newcommand{\JASRI}{Japan Synchrotron Radiation Research Institute, (JASRI), Sayo, Hyogo, 679-5198, Japan.}
\newcommand{\ENSP}{Ecole Normale Superieure de Paris, Paris, France.}
\newcommand{\NDAJ}{National Defense Academy of Japan, Yokosuka, Kanagawa 239-8686, Japan.}
\newcommand{\FHI}{Fritz Haber Institute of the Max Planck Society, Berlin, Germany.}
\begin{document}

\title{Polarization-Resolved Extreme Ultraviolet Second Harmonic Generation from LiNbO$_3$}
\linespread{1.00}
\author{Can~B.~Uzundal}
\email[Correspondence to: ]{can\_uzundal@berkeley.edu}
\affiliation{\UCBChem}
\affiliation{\LBNL}
\author{Sasawat~Jamnuch}
\affiliation{\ATLAS}
\author{Emma~Berger}
\affiliation{\UCBChem}
\affiliation{\LBNL}
\author{Clarisse~Woodahl}
\affiliation{\UoF}
\author{Paul~Manset}
\affiliation{\ENSP}
\author{Yasuyuki~Hirata}
\affiliation{\NDAJ}
\author{Toshihide~Sumi}
\affiliation{\UoT}

\author{Angelique~Amado}
\affiliation{\UCBChem}
\affiliation{\LBNL}

\author{Hisazumi~Akai}
\affiliation{\UoT}

\author{Yuya~Kubota}
\affiliation{\Spr}
\affiliation{\JASRI}

\author{Shigeki~Owada}
\affiliation{\Spr}
\affiliation{\JASRI}

\author{Kensuke~Tono}
\affiliation{\Spr}
\affiliation{\JASRI}

\author{Makina~Yabashi}
\affiliation{\Spr}
\affiliation{\JASRI}

\author{John~W.~Freeland}
\affiliation{\ANL}
\author{Craig~P.~Schwartz}
\affiliation{\LBNLC}
\author{Walter~S.~Drisdell}
\affiliation{\LBNLC}
\affiliation{\JCAP}
\author{Iwao~Matsuda}
\affiliation{\UoTQ}
\affiliation{\UoT}
\author{Tod~A.~Pascal}
\affiliation{\ATLAS}
\affiliation{\UCSDMSE}
\affiliation{\UCSDSPEC}
\author{Alfred Zong}
\affiliation{\UCBChem}
\affiliation{\LBNL}
\author{Michael~Zuerch}
\email[Correspondence to: ]{mwz@berkeley.edu}
\affiliation{\UCBChem}
\affiliation{\LBNL}
\affiliation{\FHI}

%\\date{\today}

\begin{abstract}
Second harmonic generation (SHG) spectroscopy ubiquitously enables the investigation of surface chemistry, interfacial chemistry as well as symmetry properties in solids. Polarization-resolved SHG spectroscopy in the visible to infrared regime is regularly used to investigate electronic and magnetic orders through their angular anisotropies within the crystal structure. However, the increasing complexity of novel materials and emerging phenomena hamper the interpretation of experiments solely based on the investigation of hybridized valence states. Here, polarization-resolved SHG in the extreme ultraviolet (XUV-SHG) is demonstrated for the first time, enabling element-resolved angular anisotropy investigations. In non-centrosymmetric LiNbO$_3$, elemental contributions by lithium and niobium are clearly distinguished by energy dependent XUV-SHG measurements. This element-resolved and symmetry-sensitive experiment suggests that the displacement of Li ions in LiNbO$_3$, which is known to lead to ferroelectricity, is accompanied by distortions to the Nb ion environment that breaks the inversion symmetry of the NbO$_{6}$ octahedron as well. Our simulations show that the measured second harmonic spectrum is consistent with Li ion displacements from the centrosymmetric position by $\sim$0.5~\si{\angstrom} while the Nb-O bonds are elongated/contracted by displacements of the O atoms by $\sim$0.1~\si{\angstrom}. In addition, the polarization-resolved measurement of XUV-SHG shows excellent agreement with numerical predictions based on dipole-induced SHG commonly used in the optical wavelengths. This constitutes the first verification of the dipole-based SHG model in the XUV regime. The findings of this work pave the way for future angle and time-resolved XUV-SHG studies with elemental specificity in condensed matter systems.
\end{abstract}
\maketitle
\linespread{1.50}

Nonlinear spectroscopies have become an indispensable tool to characterize material properties and dynamics. Second-order optical nonlinearities proportional to the second-order susceptibility, $\chi^{(2)}$, are especially relevant due to their distinct selection rules beyond the angular momentum selection rule. Factors such as the bulk symmetry of the crystal and the presence of an inversion center determines whether second-order susceptibilities are nonzero\cite{boyd_nonlinear_2008}. Owing to these properties, second-order nonlinear spectroscopies, such as second harmonic generation (SHG) with optical and infrared wavelengths, is widely used as an interfacial and surface probe of electronic properties of solid state materials \cite{heinz_study_1985}.

Due to the nonlinear process, a high electric field strength is needed to generate experimentally detectable SHG signals. In this regard, advances in pulsed laser sources with a high peak field strength greatly accelerated the adoption of this technique. In the optical regime, besides direct measurement of SHG, the angular anisotropy of SHG is often used to characterize electronic and magnetic orders in solids, offering an ultra-sensitive probe of crystalline symmetry \cite{torchinsky_rotational_2017}. For example, SHG angular anisotropy has been used to characterize the symmetries of ferroic materials \cite{denev_magnetic_2008,padmanabhan_linear_2018}, multipolar order \cite{Jin2019}, and chiral structures \cite{fiebig_second-harmonic_2005,fichera_second_2020}. In systems with strong electron correlations -- from unconventional superconductors to quantum spin liquid -- the sensitivity afforded by SHG has also revealed important phases that elude previous investigations \cite{Harter2017,Zhao2017,Laurita2019}.
Free-electron lasers (FELs) present a similar opportunity to extend the capabilities of SHG into the extreme ultraviolet (XUV) and soft X-ray regime \cite{lam_soft_2018}. Recently, nonlinear X-ray and XUV spectroscopies have been the subjects of both theoretical \cite{minerbi_difference_2019} and experimental \cite{glover_x-ray_2012,szlachetko_establishing_2016,beye_non-linear_2019,bohinc_nonlinear_2019,shwartz_x-ray_2014} works as a result of the available high intensity light sources. Among these techniques, XUV-SHG is particularly attractive as the core-level specificity of XUV radiation and the unique selection rules of SHG can be united. Further, short pulse durations pushing to the attosecond regime that are available at the current and upcoming generation of FELs is expected to enable XUV-SHG studies with exceptional time resolution \cite{duris_tunable_2020}. Experimentally, distinct elemental edges can be probed using XUV-SHG \cite{lam_soft_2018, yamamoto_element_2018}. Following these proof of principle works, XUV-SHG was also shown to probe material properties with high sensitivity to the chemical environment around the select elements \cite{berger_direct_nodate} and was demonstrated as a spectroscopic tool capable of probing buried interfaces \cite{schwartz_angstrom-resolved_nodate}.

In contrast to optical SHG that measures an average response across the elements forming the valence orbitals, XUV-SHG probes the element-specific core level states, allowing the separation of elemental contributions in the measured anisotropies. The elemental specificity of XUV-SHG is particularly attractive for materials where emergent behavior is rooted in single ion displacements in the unit cell. For instance, in ferroelectric materials, a spontaneous polarization forms as a result of unit cell distortions that break the inversion symmetry. In particular for the ferroelectric material studied in this article, LiNbO$_3$, the spontaneous polarization establishes as a result of Li displacement relative to the Nb-O octahedron in the unit cell.

Here, polarization-resolved XUV-SHG is demonstrated for the first time and applied to study the nature of symmetry-breaking ion displacement in ferroelectric LiNbO$_{3}$. Spectroscopy measurements covering the Li $K$ and Nb $N$ edges were conducted in concert with polarization-resolved studies at an FEL. Resonant features relating to the two elements are observed and assigned using \textit{ab initio} density functional perturbation theory (DFPT). The angular anisotropy of a selected resonance is resolved which is well reproduced by the theory of nonlinear polarization based on  the DFPT calculation.

The experiments were performed at the BL1 of SACLA in Japan \cite{owada_soft_2018}. A $\sim$30-fs $p$-polarized FEL pulse was tuned to energies between 28~eV and 33~eV with 0.5~eV steps and incident on an $x$-cut LiNbO$_3$ crystal (Fig.~S5) at 45$^\circ$ with respect to the sample surface. The incoming photon energies are referred as the \textit{fundamental} in the remaining text. The second harmonic response of the sample was analyzed in two separate experiments. 

\begin{figure*}[p]\centering
    \includegraphics[width=1\textwidth]{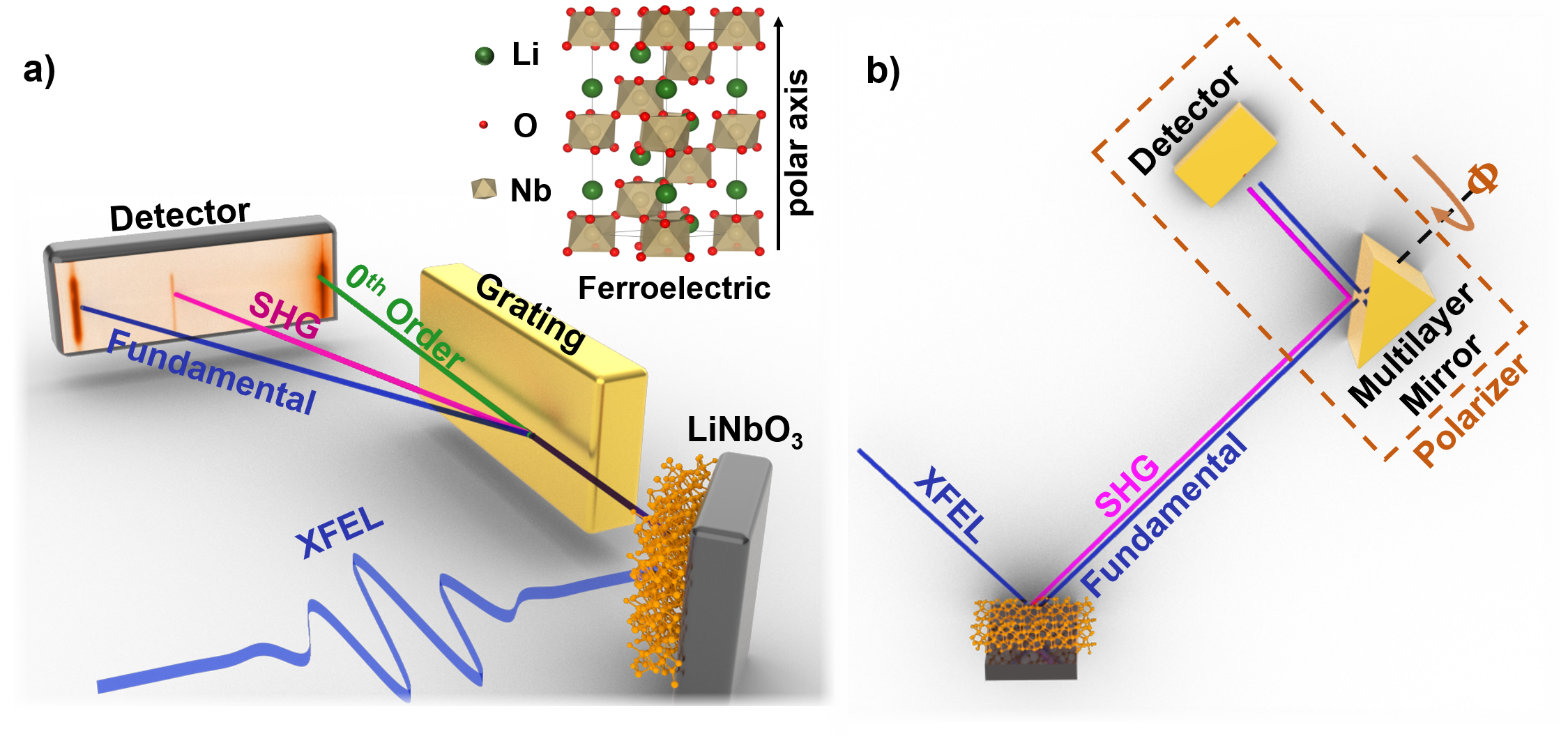}
    \caption {Schematic overviews of the two experiments. In both experiments the $p$-polarized fundamental is incident on an $x$-cut LiNbO$_3$ sample at \SI{45}{\degree} with respect to the crystal surface. \SI{45}{\degree} incident angle is the Brewster's angle for XUV energies. SHG is also detected at \SI{45}{\degree} with respect to the crystal surface. a) Schematic illustration of the spectral measurement. The FEL beam is incident on the LiNbO$_3$ in a reflection geometry. The residual fundamental and the emitted second harmonic of the incident FEL are analyzed by dispersing with a grating and imaging with an MCP detector. Inset shows the unit cell of LiNbO$_3$ in its ferroelectric state. b) Schematic illustration of the second harmonic polarization measurement. The residual fundamental and the emitted second harmonic are reflected off of a multilayer mirror tuned to 66~eV. The polarization of the residual fundamental and the second harmonic are simultaneously measured.}
    \label{fig:intro}
\end{figure*}

In the first experiment, as illustrated in Fig.~\ref{fig:intro}(a), the second harmonic and the reflected fundamental were dispersed by a grating (1200~groove/mm, 30-002, Shimadzu) and captured using a microchannel plate detector (MCP) (Rectangular, Hamamatsu Photonics) coated with CsI. The images of the detector were captured with a camera (IPX-VGA120-LMCN, Imperx Inc.). This measurement was used to retrieve the second order susceptibility spectrum, $\chi^{(2)}_{\text{eff}}(\omega)$. We emphasize that both the fundamental and the second harmonic light were simultaneously recorded on the detector, allowing a comparison of the shot-to-shot variation in the photon flux and energy. The photon energy jitter was approximately 0.2\% of the fundamental frequency and  was leveraged to increase the spectral resolution (see Supplementary Material Section S1). Shot-to-shot fluctuations in the photon flux of the fundamental was used to extract the second-order susceptibility [Fig.~S2(b)].

In the second experiment [Fig.~\ref{fig:intro}(b)], the polarization of the second harmonic was investigated with an XUV polarizer. Specifically, the second harmonic emitted from the sample was reflected off a multilayer mirror at Brewster's angle, such that only the $s$-polarized portion of the incident light was reflected. The reflected light was then detected by an MCP. The multilayer mirror and the MCP were rotated azimuthally around the beam axis [$\phi$ in Fig.~\ref{fig:intro}(b)] between $-15^\circ$ to $115^\circ$ in 3$^\circ$ steps with respect to the laboratory frame. As the multilayer mirror was rotated, the portion of light $p$-polarized with respect to the mirror surface with the new orientation is refracted into the mirror substrate and absorbed, while the $s$-polarized component is reflected. The multilayer mirror was coated such that the second harmonic is preferentially reflected over the fundamental, making the separation of the two possible \cite{attwood_soft_1999}. Nonetheless, a small fraction of fundamental was able to reach the MCP due to non-perfect extinction on the multilayer mirror. Inspecting Fig.~S3, one can estimate that the fundamental and second harmonic intensities are approximately the same order of magnitude. As a result of the residual fundamental at the detector, the polarization of the fundamental and the second harmonic can be resolved simultaneously. 

A detailed analysis procedure for the spectral data is presented in the Section S1 of the Supplementary Material. Briefly, each FEL shot imprints a 2D image on the MCP detector. Each image contains peaks associated with the incident fundamental and the emitted second harmonic at their respective frequencies. The quality of each shot was determined by fitting a Gaussian to the fundamental peak and assessing the quality of the fit. Approximately 5\% of all shots at each energy were discarded on the basis of R$^2$ values less than $0.9$. The remaining shots were binned with respect to the fundamental energy and intensity. The second order susceptibility at each energy was extracted using the quadratic relationship between the fundamental intensity and its second harmonic.

The detector in the polarization-resolved experiment measures both the reflected fundamental and the emitted second harmonic as a function of angle $\phi$ as shown in Fig.~\ref{fig:intro}(b). A detailed procedure for the data analysis steps for the polarization experiment is presented in the Supplementary Material Section~S2. Briefly, the fundamental and the second harmonic response were separated by a linear background subtraction. At lower incident fundamental intensities, the measured voltage was linear with respect to the intensity of the incident fundamental while at higher incident fundamental intensities a quadratic relationship is observed. 

These experiments were corroborated by theoretical calculations. The linear response of LiNbO$_3$ was simulated with first-principles density functional theory \cite{hohenberg_inhomogeneous_1964} using the \textbf{exciting}\cite{gulans_exciting_2014} full potential all electron augmented linearized planewave package. Both centrosymmetric and noncentrosymmetric structures of LiNbO$_3$ were investigated. The Brillouin zone was sampled with a $15\times15\times15$ $\Gamma$-point centered $k$-point grid within the local density approximation \cite{perdew_self-interaction_1981}. The Li core 1$s$ and Nb semi-core 4$s$ and 4$p$ electrons were included in the self-consistent field calculation
loop to extract their respective linear responses. The excited states of the system were accessed through time-dependent density functional theory (TDDFT) simulations using the random phase approximation (RPA) kernel implemented within exciting\cite{vast_local_2002,sagmeister_time-dependent_2009} with a $q$-point chosen to be the same as that of the aforementioned  k-point grid. Plots of imaginary part of dielectric function are shown in Fig.~S12. No characteristic differences to the linear dielectric response were found between the non-polar and polar phases. The experimental and theoretical results show good agreement except the overestimation of the first peak at 36~eV as shown in Fig.~S12, which is a common trait in the level of theory used here. To properly sample the $t_{2g}$ peak, many-body effects would need to be included, but is beyond the scope of the present work and not necessary for the current level of analysis. The level of theory employed in the linear response calculation was kept at the same level as the second harmonic response calculation for consistency and ease of comparison. The second harmonic response formalism by Sharma\cite{sharma_second-harmonic_2004} implemented within \textbf{exciting}\cite{gulans_exciting_2014} was used to calculate the second order susceptibility of LiNbO$_{3}$. Here, lifetime broadening was also employed to account for the high oscillatory behavior of high-energy states as detailed in the work of Lam\cite{lam_soft_2018}. A total of 120 empty states were included to account for excitation up to double of the incoming photon frequency. Molecular dynamics simulations were performed to correctly reflect the finite temperature of the system (see SI Section S4 for details).

\begin{figure*}[p]
    \centering
    \includegraphics[width=1\textwidth]{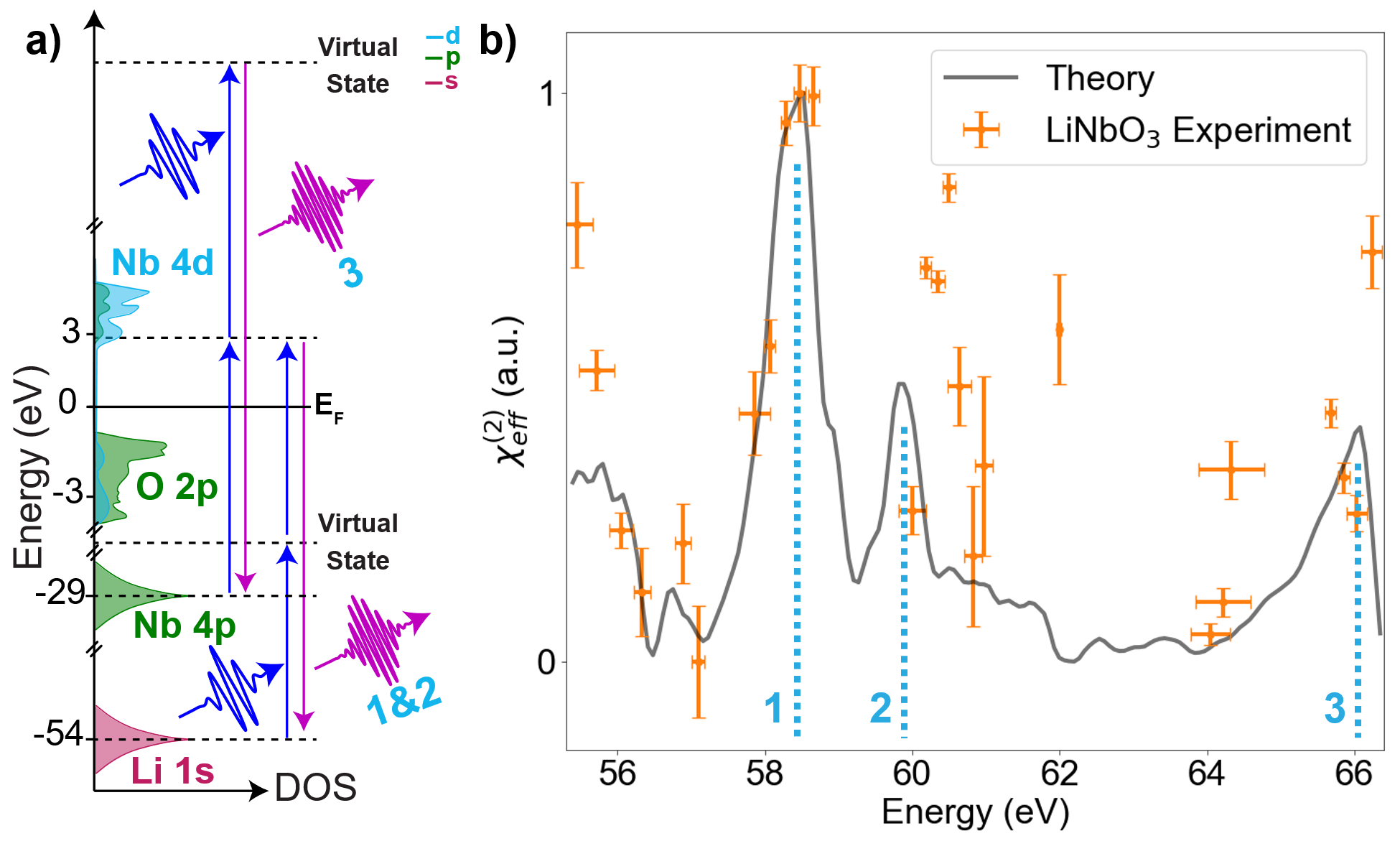}
    \caption {Resonant transitions for LiNbO$_3$ within the range of fundamental energies studied in the spectral experiment. (a)~Schematic illustration of the possible transitions with respect to the electronic density of states. Two prominent transitions that involve the Li 1$s$ and Nb 4$p$ core states are labeled. For the Li 1$s$ states (Transitions~1 and 2), SHG occurs through a half resonant scheme and it is facilitated by a virtual state, while the transition originating from Nb 4$p$ (Transition~3) is facilitated via the conduction band states of majority Nb 4$d$ character. (b)~Measured and theoretical values of the effective $\chi^{(2)}_{\text{eff}}(\omega)$ spectrum for LiNbO$_3$ showing three prominent features corresponding to the labeled transitions from panel (a).}
    \label{fig:transition}
\end{figure*}

Considering the electronic density of states, the core level transitions that can be accessed by XUV-SHG is schematically shown in Fig.~\ref{fig:transition}(a). Half resonant transitions from Li 1$s$ core states to the conduction band states with majority Nb 4$d$ character, along with resonant transitions originating from Nb 4$p$ core-like states fall within the range of the fundamental energies studied. Though the availability of density of states is vital for observing XUV-SHG, it is not the only factor that governs the transition probability. Experimentally, the measured second order susceptibility spectrum, $\chi^{(2)}_{\text{eff}}$, is a direct measurement of the allowed transitions within the selection rules of XUV-SHG. In Fig.~\ref{fig:transition}(b) the measured $\chi^{(2)}_{\text{eff}}$ is overlaid with the theoretically calculated spectrum for LiNbO$_3$. The theoretically calculated spectrum is derived as a weighted sum of the individual $\chi^{(2)}$ tensor elements of the C$_{3v}$ point group under consideration of the experimental geometry (see SI Section S3 for details). The two half-resonant features around 58~eV correspond to transitions from Li 1$s$ to conduction band states of majority Nb 4$d$ character. These features report on the ferroelectric displacement in the unit cell involving the Li ions. Our TDDFT calculations (SI Section S4) are consistent with an inversion symmetry breaking displacement of Li ions by  $\sim$.5~\si{\angstrom}. The resonant feature at 58~eV is similar to the feature observed in a previous XUV-SHG study on LiOsO$_3$\cite{berger_direct_nodate}. The previous study reported an enhancement of the $\chi^{(2)}$ amplitude with inversion symmetry breaking in the unit cell by Li displacements. In fact, the behavior of the two Li containing compounds with ferroelectric displacements, LiNbO$_3$ and LiOsO$_3$ is very similar at this energy range. The key difference in the $\chi^{(2)}_{\text{eff}}$ spectra  is highlighted in Fig.~\ref{fig:transition}(b) as transition \#3. This highlighted feature in Fig.~\ref{fig:transition}(b) stems from half-resonant transitions from the Nb $N$ edge, pointing towards inversion symmetry breaking not only around the Li ion but also the Nb ion. Molecular dynamics simulations of the Nb ion environment in the presence of the ferroelectric displacement show that the amplitude of $\chi^{(2)}_{\text{eff}}$ at $2\hbar\omega=66$~eV is consistent with a modulation of the Nb-O bonds by $\sim$0.1\si{\angstrom}. The molecular dynamics simulations also show that antiferrodistortive rotations of the NbO${_6}$ does not substantially contribute to the amplitude of the $\chi^{(2)}_{\text{eff}}$ spectrum shown in Fig.~\ref{fig:transition}(b). This result is consistent with a picture of ferroelectric displacement of the Li ion that is stabilized by distortions to the NbO$_6$ octahedra that break the inversion symmetry around the Nb ion in the unit cell \cite{toyoura_first-principles_2015,barker_dielectric_1967,inbar_comparison_1996,inbar_origin_1997}. No evidence of such inversion symmetry breaking around the Os ion in LiOsO$_3$ was observed \cite{berger_direct_nodate}. This contrast between the two cases demonstrates the strength of the unique selection rules of XUV-SHG and is a direct visualization of element specificity of the method.

\begin{figure*}[p]
    \centering
    \includegraphics[width=0.8\textwidth]{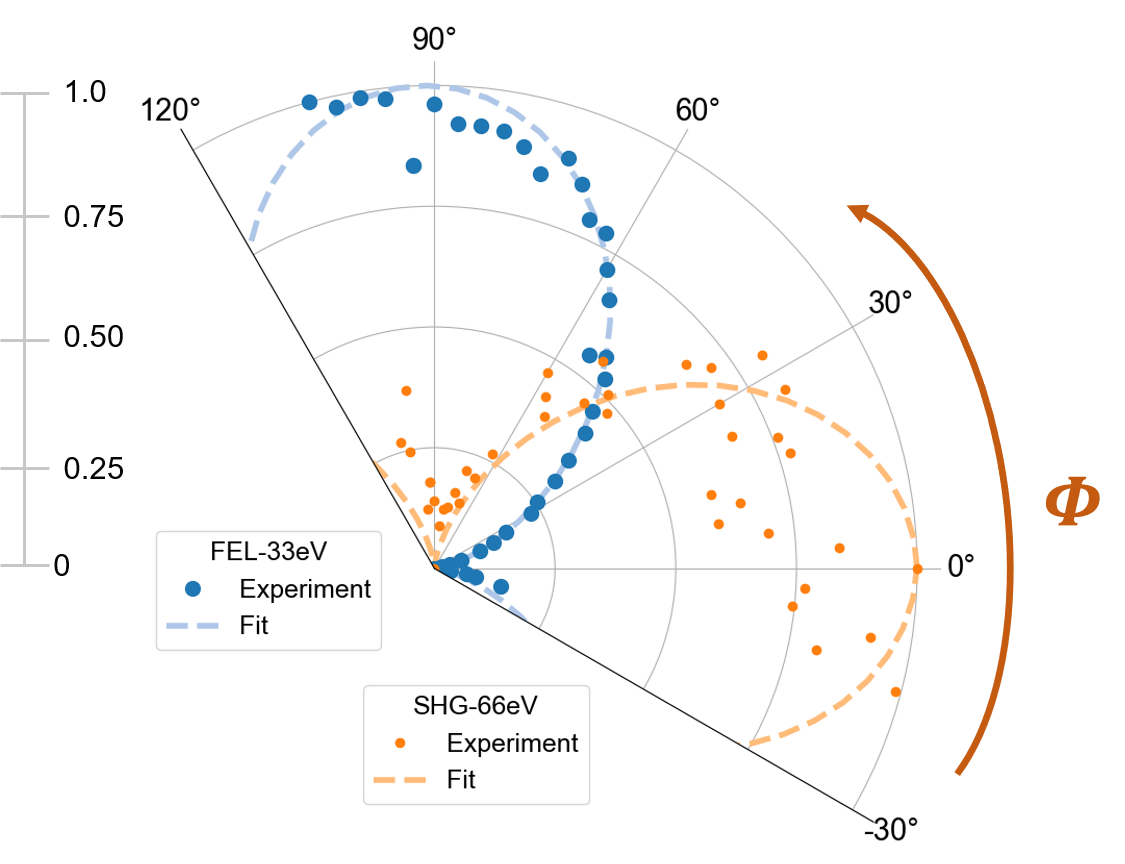}
    \caption {Results of the polarization-resolved experiment on $x$-cut LiNbO$_3$. Measured polarization of the incident FEL (blue dots) at 33~eV with the characteristic Malus' Law fit (blue dashed lines) and the polarization of the SHG (orange dots) at 66~eV with the fit to the Malus' Law (orange dashed lines). $p$-polarization with respect to the laboratory frame is \SI{90}{\degree} while $s$-polarization is \SI{0}{\degree}. Scale bar shows the amplitude of the polarization in arbitrary units.}
    \label{fig:polar}
\end{figure*}

To further demonstrate the capabilities of energy-resolved SHG, we conducted polarization-resolved measurements at a single energy. Along with the fundamental ($\hbar\omega=33$~eV) that initiates SHG, the polarization of SHG at $2\hbar\omega=66$~eV is resolved and shown as a polar plot in Fig.~\ref{fig:polar}. The expected $p$-polarization of the fundamental is recovered while the second harmonic is measured to be $s$-polarized. Using the well established arguments for the crystal symmetry and a detailed tensor analysis, the polarization of the second harmonic can be calculated and decomposed into four distinct channels characterized by the polarization of the incident fundamental and the detection channel of the polarization. All four cases are shown in Fig.~S8 as a function of in plane rotation of $x$-cut and $z$-cut LiNbO$_3$. The calculated polar plots for the 4 distinct channels of polarization ($p$ in $p$ out, $p$ in $s$ out, $s$ in $p$ out, $s$ in $s$ out) show distinct patterns demonstrating the symmetry of the crystal. Note that, as expected, the 3-fold symmetry of the bulk crystal is reproduced by the polar plots for the $z$-cut of the crystal. For $x$-cut LiNbO$_3$, when it is incident with a $p$-polarized fundamental, the majority of the emitted second harmonic is calculated to be $s$-polarized [Fig.~S7(b) \& (c)] at $2\hbar\omega=66$~eV, in excellent agreement with the measured polarization of the second harmonic. The emitted $s$-polarized SHG constitutes the emission of the orthogonal polarization to the polarization of the fundamental. The emission of the orthogonal polarization can be attributed to the energy dependent $\chi^{(2)}$ tensor as the emission of $p$-polarized SHG is indeed symmetry allowed as shown in Fig.~S7(b) \& (c) for a different XUV energy. The observed $s$-polarized SHG, thus cannot be solely explained by symmetry considerations and highlights the nonlinear nature of the light-matter interaction in this regime. 

The findings presented here demonstrate the feasibility of an XUV-SHG study with angular resolution and the possibility to extend SHG rotational anisotropy studies into XUV wavelengths. The long established theory of nonlinear polarization up to the dipole contributions is shown to be adequate to explain the measured polarization for a bulk noncentrosymmetric material in the extreme ultraviolet. Our spectral findings suggest that inversion symmetry breaking in the ferroelectric phase of LiNbO$_{3}$ may not be limited to Li ion displacements. Contributions to the $\chi^{(2)}_{\text{eff}}$ spectrum from Nb ions suggest inversion symmetry breaking around the Nb ions is important as well. We envision that the demonstrated principles can be fully extended to time-resolved studies that leverage the excellent time resolution afforded by the short XUV pulses, paving the way towards attosecond nonlinear spectroscopies on surfaces or buried interfaces with element specificity.

\section{ACKNOWLEDGEMENTS}
\noindent M. Z., C. S. and A. A. acknowledge support by the Max Planck Society (Max Planck Research Group). M. Z. acknowledges support by the Federal Ministry of Education and Research (BMBF) under ``Make  our  Planet  Great  Again – German  Research  Initiative''  (Grant  No.~57427209 ``QUESTforENERGY'') implemented by DAAD. J. F. acknowledges Department of Energy grant number DE SC-0012375. W.D. acknowledges support from the Joint Center for Artificial Photosynthesis, a DOE Energy Innovation Hub, supported through the Office of Science of the U.S. Department of Energy, under Award No. DE-SC0004993. A.Z. acknowledges support from the Miller Institute for Basic Research in Science. Measurements were performed at BL1 of SACLA with the approval of the Japan Synchrotron Radiation Research Institute (JASRI) (Proposal No. 2019B8066). This work was supported by the SACLA Basic Development Program 2018-2020. The authors would like to acknowledge the supporting members of the SACLA facility. Additional measurements were performed at beamline 6.3.2 of the Advanced Light Source, a U.S. DOE Office of Science  User Facility under contract no. DE-AC02-05CH11231. This research used resources of the National Energy Research Scientific Computing Center, a DOE Office of Science User Facility supported by the Office of Science of the U.S. Department of Energy under Contract No. DE-AC02-05CH11231. This work also used the Extreme Science and Engineering Discovery Environment (XSEDE), which is supported by National Science Foundation grant number ACI-1548562. C. W. acknowledges support by the National Science Foundation REU Program grant number 1852537. M. Z. acknowledges funding by the W. M. Keck Foundation, funding from the UC Office of the President within the Multicampus Research Programs and Initiatives (M21PL3263), and funding from Laboratory Directed Research and Development Program at Berkeley Lab (107573). We acknowledge Shukai Yu for providing the sample for this study. C. U. is grateful for discussions with Yue Sun.

\clearpage
\bibliography{refs}

\end{document}